\documentclass[conference,a4paper]{IEEEtran}
\usepackage{amssymb, amsthm,bm} 
\usepackage{color}
\usepackage{comment}
\usepackage[utf8]{inputenc} 
\usepackage[T1]{fontenc}
\usepackage{url}             
\usepackage{cite}             

\usepackage[cmex10]{amsmath} 
\usepackage{mleftright}       
\mleftright                   

\newtheorem{theorem}{Theorem}
\newtheorem{problem}{Problem}
\newtheorem{lemma}[theorem]{Lemma}
\newtheorem{corollary}[theorem]{Corollary}

\usepackage{graphicx}         
\usepackage{siunitx, booktabs}         
                              
\usepackage{tikz}
\usetikzlibrary{arrows,automata, positioning}
\usetikzlibrary{calc, chains,
                fit,
                positioning,
                shapes}
\usetikzlibrary{patterns}     
\usepackage{pgfplots}                        
                              
\newcommand{\hbin}[1]{\ensuremath{h\mleft(#1\mright)}}
\newcommand{\trin}[2]{\ensuremath{g\mleft(#1,#2\mright)}} 
\newcommand{\Ls}{\ensuremath{\Sigma}}
\newcommand{\Repr}{\ensuremath{U}} 
\newcommand{\Pwt}{\ensuremath{V}} 
\newcommand{\Ewt}{\ensuremath{E}} 
\newcommand{\Rset}{\ensuremath{\mathbb{E}}}
\renewcommand{\vec}[1]{\ensuremath{\bm{#1}}}      
\newcommand{\mat}[1]{\ensuremath{\bm{#1}}}

\begin{document}

\title{Generic Decoding of Restricted Errors}

\author{%
  \IEEEauthorblockN{Marco Baldi$^1$, Sebastian Bitzer$^2$, Alessio Pavoni$^1$, Paolo Santini$^1$, Antonia Wachter-Zeh$^2$, Violetta Weger$^2$}
  \IEEEauthorblockA{$^1$Marche Polytechnic University, Italy\\
                    $^2$Technical University of Munich, Germany}
}
\maketitle

\begin{abstract}
Several recently proposed code-based cryptosystems base their
security on a slightly generalized version of the classical (syndrome) decoding problem. 
Namely, in the so-called restricted (syndrome) decoding problem, the error values stem from a restricted set.
In this paper, we propose new generic decoders, that are inspired by subset sum solvers and tailored to the new setting.
The introduced algorithms take the restricted structure of the error set into account in order to utilize the representation technique efficiently.
This leads to a considerable decrease in the security levels of recently published code-based cryptosystems. 
\end{abstract}

\section{Introduction}\label{sec:intro}
With the recent advances in quantum technology, the search for quantum-secure  cryptographic systems has become one of the most pressing challenges. In the NIST selection process of post-quantum cryptosystems launched in 2016, which has now reached the 4th round, some of the most promising candidates are based on algebraic coding theory, more precisely on the hardness of decoding a random linear code. 
A new trend in code-based cryptography is to base the security on the hardness of a slightly different problem, e.g. \cite{wave}.  
One such new decoding problem is a generalized decoding problem, where  one restricts the error set. Some recent systems base their security on this decoding problem \cite{oldrest, freudenberger}.  

In this paper, we provide new solvers for such  settings, which are inspired by subset sum solvers \cite{bcj}. The idea is to use the additive structure that can be found in the error set and to add only few elements to the search set, such that one achieves  more representations of the elements in the error set and does not increase the search sizes too much. The connection between subset sum solvers and generic decoders has often been exploited (see, e.g., \cite{hgj, bjmm}). We show the impact of these new attacks on systems to which they apply and how significantly their security levels decrease. As the restricted decoding problem is still very promising for cryptographic applications, the presented attacks  should be considered in future proposals. For this purpose, we also shortly explain how the specific solvers can be generalized to any setting.  
  
  The paper is structured as follows: 
  in Section \ref{sec:prel}, we recall some basic notions of algebraic coding theory and introduce the required notation. 
  In Section \ref{sec:restsdp}, we introduce the restricted decoding problem and discuss some of its properties. 
  We then present the new attacks in Section \ref{sec:solvers}, comparing the approaches in different cases and computing new security levels for cryptosystems that are using the restricted decoding problem. Finally, Section \ref{sec:conclusion} concludes the paper.

\section{Preliminaries}\label{sec:prel}
Throughout this paper we denote by $q$  a prime power and by $\mathbb{F}_q$ a finite field of order $q$. We denote  the identity matrix of size $n$ by $\mat{\text{Id}}_n.$  
For a set $J$, we denote by $J_0=J\cup\{0\}$. 
For $ x\in [0,1]$, we denote by $h(x)$ the binary entropy function.  
For $n \geq k_1+k_2$ we denote by $\binom{n}{k_1,k_2} =  \binom{n}{k_1+k_2}\cdot\binom{k_1+k_2}{k_1}$ the trinomial coefficient.
Recall that 
\begin{align*}
    \lim\limits_{n \to \infty} \frac{1}{n} \log_2\left(\binom{f_1(n)}{f_2(n)}\right)  &= F_1 \hbin{\frac{F_2}{F_1}}, \\
    \lim\limits_{n \to \infty} \frac{1}{n} \log_2\left( \binom{f_1(n)}{f_2(n),f_3(n)}\right)
    &= F_1 g\mleft(\frac{F_2}{F_1},\frac{F_3}{F_1}\mright),
\end{align*}  
with $F_i=\lim\limits_{n \to \infty}\frac{f_i(n)}{n}$, $\hbin{x}$ the binary entropy function and
\[
g(x,y) = -x\log_2(x) - y\log_2(y) - (1-x-y) \log_2(1-x-y).
\]

A \emph{linear code} $\mathcal{C}$ is a $k$-dimensional subspace of $\mathbb{F}_q^n$.
A linear code can be compactly represented either through a \emph{generator matrix} $\mat G \in \mathbb{F}_q^{k \times n}$ or through a \emph{parity-check matrix} $\mat H \in \mathbb{F}_q^{(n-k) \times n}$, which have the code as image or as kernel, respectively. We say that a linear code has \emph{rate} $R=k/n.$ 
We define $\mathcal{C}_J$ as  $\mathcal{C}_J= \{\vec c_J \mid \vec c \in \mathcal{C}\}, $ where $\vec{c}_J$ is the projection of $c$ on the coordinates indexed by $J.$
For any $\vec x \in \mathbb{F}_q^n$, we call $\vec s= \vec x \mat H^\top \in \mathbb{F}_q^{n-k}$ a \emph{syndrome}. 
A set $I \subseteq \{1, \ldots, n\}$ of size $k$ is called an \emph{information set} for $\mathcal{C}$, if $\left\lvert \mathcal{C}\right\rvert = \left\lvert \mathcal{C}_I \right\rvert.$
We say that a generator matrix, respectively a parity-check matrix, is in systematic form (with respect to the information set $I$), if the columns of $\mat G$ indexed by $I$ form $\mat{\text{Id}}_k$, respectively, if the columns of $H$ not indexed by $I$ form $\mat{\text{Id}}_{n-k}.$
We endow the vector space $\mathbb{F}_q^n$ with the \emph{Hamming metric}: 
the Hamming weight of a vector $\vec x \in \mathbb{F}_q^n$ is given by the number of its non-zero entries, i.e., \vspace{-0.4em}
\[ \text{wt}_H(\vec x) = \left\lvert \{i \in \{1, \ldots, n \} \mid x_i \neq 0 \} \right\rvert,\]
which then induces a distance, as $d_H(\vec x,\vec y)=\text{wt}_H(\vec x- \vec y)$, for $\vec x,\vec y \in \mathbb{F}_q^n.$

\section{Restricted Syndrome Decoding Problem}\label{sec:restsdp}
Throughout this paper, we consider the computational version of the following decisional problem. 

\begin{problem}[Restricted Syndrome Decoding Problem (RSDP)]
Let $g \in \mathbb{F}_q$ be an element of order $z$ and define the error set $\mathbb{E}= \{g^i \mid i \in \{1, \ldots, z\}\}$. Let $\mat H \in \mathbb{F}_q^{(n-k) \times n}, w\in \mathbb{N}$ and $\vec s \in \mathbb{F}_q^{n-k}.$ Does there exist a vector $\vec e \in \mathbb{E}_0^n = (\mathbb{E}\cup\{0\})^n$ with $\text{wt}_H(\vec e)=w$, such that $\vec s=\vec e\mat H^\top?$
\end{problem}

If we choose $g$ to be a primitive element of $\mathbb{F}_q$ we can indeed recover the original syndrome decoding problem (SDP) in the Hamming metric.
Thus, the problem directly inherits the NP-completeness.
Even more is true, for a fixed  $g$, the RSDP is still NP-complete,  this follows directly from \cite[Proposition~2]{LeeNP}.  
 \begin{corollary}
The RSDP with a fixed $\mathbb E$ is NP-complete.
 \end{corollary} 
In addition to generalizing the classical SDP, the RSDP also covers the case 
$\mathbb{E}= \{\pm 1\}$ considered in \cite{oldrest}, by setting  $g=-1$, respectively $z=2$. 

Also \cite{freudenberger} considers a particular case of the RSDP. 
In their paper, they consider $\mathbb{F}_q$, where $q=6m+1$ is a prime, that is the field allows for an element of order $6$. They then consider the errors to live in $\mathbb{E}= \{ \pm 1, \pm g, \pm(g-1)\}.$
This corresponds to the RSDP with $z=6$. 
In fact, since
$g$ is a root of $x^6-1=(x^3-1)(x+1)(x^2-x+1)$ and $g$ is not a root of $(x^3-1)$ or $(x+1)$, we must have $g^2=g-1.$
 
From the uniqueness condition $\binom{n}{w}z^wq^{k-n}\leq 1$, we can easily observe  that the restriction on the entries of the error vector $\vec e$ allows us to increase the Hamming weight of $\vec e$, while still having a single solution to the RSDP with high probability. 
This is the main motivation for introducing the RSDP in cryptographic applications.

\section{Solvers for the RSDP}\label{sec:solvers}
There exist several algorithms that could potentially be applied to the RSDP, for example statistical decoders, such as~\cite{stat2}, or attacks from lattice-based cryptography, such as \cite{latt}.
In this paper, we focus on Information Set Decoding (ISD) algorithms combined with ideas from subset sum solvers, such as \cite{bcj}. 

The history of ISD dates back to the algorithm of Prange~\cite{prange1962use} in 1962 and has resulted in several improvements (for an overview in the binary case see \cite{meurer,survey}). 
Variants of Prange's ISD with smaller computational complexity are Stern/Dumer \cite{stern, dumer}, MMT \cite{mmt} and BJMM \cite{bjmm, bjmmFq}. We only shortly recall their ideas in the following, before we adapt them to our setting.  
Given $\mat H \in \mathbb{F}_q^{(n-k) \times n}, w \in \mathbb{N}$ and $\vec s \in \mathbb{F}_q^{n-k}$, one starts by choosing a set $J \subseteq \{1, \ldots, n\}$ of size $k+\ell$, for some positive integer $\ell \leq n-k$, which contains an information set. 
One then brings the parity-check matrix into systematic form, by performing Partial Gaussian Elimination (PGE), denoted by $\mat{H}'$, and performs the same operations on the syndrome. 
For simplicity, assume that the set $J$ is chosen in the first $k+\ell$ positions. Thus, we get the syndrome equations
$$ \vec e\mat{H}'^\top = \begin{pmatrix} \vec{e}_1 & \vec{e}_2 \end{pmatrix} \begin{pmatrix} \mat{\text{Id}}_{n-k-\ell} & \mat 0 \\  \mat{A}_1  & \mat{A}_2 \end{pmatrix} = \begin{pmatrix} \vec{s}_1 & \vec{s}_2 \end{pmatrix},$$ where $\mat{A}_1 \in \mathbb{F}_q^{(k+\ell) \times (n-k-\ell)}, \mat{A}_2 \in \mathbb{F}_q^{ (k+\ell) \times \ell}, \vec{e}_1 \in \mathbb{E}_0^{n-k-\ell}, \vec{e}_2 \in \mathbb{E}_0^{k+\ell}, \vec{s}_1 \in \mathbb{F}_q^{n-k-\ell}$ and $\vec{s}_2 \in \mathbb{F}_q^\ell$. 
One first solves for $\vec{e}_2$, assuming a weight $v$, i.e., $\vec{e}_2\mat{A}_2^\top=\vec{s}_2$ and then checks if $\vec{e}_1=\vec{s}_1-\vec{e}_2\mat{A}_1^\top$ has the remaining weight $w-v$ and entries in $\mathbb{E}_0$.
To solve the smaller instance given by $(\mat{A}_2,v,\vec{s}_2)$ one can use different approaches. 

Before we start describing these approaches, let us fix some notation. 
To compare algorithms for fixed  $Q = \log_2(q)$,
$Z = \log_2(z)$, we are interested in the asymptotic cost, that is, we write the cost as $2^{F(R,W)n}$, for some function $F(R,W)$, where  $R= \lim\limits_{n \to \infty} \frac{k(n)}{n}, W=\lim\limits_{n \to \infty} \frac{w(n)}{n}$.  
Since we have seen that it is enough to solve the smaller instance of weight $v$ and length $k+\ell$,  we also write $L=\lim\limits_{n \to \infty} \frac{\ell(n)}{n} \leq 1-R, V=\lim\limits_{n \to \infty} \frac{v(n)}{n} \leq \min\{W, R+L\}$, which are internal parameters and can hence be optimized.
Then, the complexity of a decoder using the PGE setup is given by the following theorem, see, e.g., \cite{meurer}.
\begin{theorem}
A generic decoder using PGE has time complexity $2^{F(R,W)n}$, with 
$F(R,W)= N(R,W, L,V) + C(R,L,V)$, where $N(R,W,L,V)$ denotes the number of iterations, i.e.,
$$  \hbin{W}-(R+L)\hbin{\tfrac{V}{R+L}}-(1-R-L)\hbin{\tfrac{W-V}{1-R-L}}
$$  
and $C(R,L,V)$ denotes the time complexity of solving the smaller instance, i.e., the time to enumerate all solutions of the smaller instance under the assumed weight distribution.
\end{theorem}
We  compare our algorithms to  different approaches, namely Stern/Dumer \cite{stern, dumer} and BJMM \cite{bjmm, bjmmFq}, which encompasses MMT as well \cite{mmt}, adapted to the new setting. 
We proceed by explaining how this adaption works.

A classical approach to enumerating all solutions of the small instance is performing a collision search.
This technique was applied to hard knapsacks by Schroeppel and Shamir \cite{schroeppelshamir} 
and adopted for the syndrome decoding problem by Stern and Dumer \cite{stern, dumer}.
In this approach one uses a  set partition of $\vec{e}_2$ into $\vec{e}_2=(\vec{x}_1,\vec{x}_2)$, where both $\vec{x}_i \in \mathbb{E}_0^{(k+\ell)/2}$ have weight $v/2$. One constructs lists containing such $\vec{x}_i$ and by a collision search finds candidates $\vec{e}_2$. 
This merging process is called concatenation merge. 
For more details on the classical algorithm we refer to \cite{stern, dumer} and for the adaption to \cite{oldrest2}.

\begin{lemma}
The enumeration cost for the smaller instance of the restricted Stern/Dumer algorithm is given by
$
C(R,L,V) = \max\{\frac{\Ls}{2}, \Ls - Q\cdot L\},
$
where
\[\Ls = (R+L)\hbin{\tfrac{V}{R+L}}+Z V\] 
is the asymptotic size of the search space, i.e., the set of all vectors that are well-formed, i.e., they satisfy the constraint under which the solutions of the small instance are enumerated.
\end{lemma}
As this is a well-known algorithm with the only change that the lists are taken in $\mathbb{E}_0$, rather than in $\mathbb{F}_q$, we  omit the proof.
 
More in general, one can perform a concatenation merge of two lists $\mathcal{L}_1, \mathcal{L}_2$, of asymptotic size $\Lambda= \lim\limits_{n \to \infty}\frac{ \log_2(\mid \mathcal{L}_i\mid)}{n}$, requiring only that they satisfy a syndrome equation on $u$ positions, with $U= Q\cdot \lim\limits_{n \to \infty} \frac{u(n)}{n}$. This costs asymptotically 
\[ 2^{n\max\{\Lambda, 2\Lambda-U\}}.\]

An alternative to this collision search is using the representation technique, which has proven efficient in solving the hard knapsacks \cite{hgj, bcj} and the classical SDP \cite{mmt, bjmm}.

Instead of a  set partition, a sum partition is used: $\vec{e}_2= \vec{x}_1+\vec{x}_2$, where in the classical case the $\vec{x}_i \in \mathbb{F}_q^{k+\ell}$  have weight $v/2+\varepsilon$. This is chosen such that   $\varepsilon$ positions of their supports are overlapping and cancel out. 
Let us first introduce the number of ways we can write $\vec{e}_2= \vec{x}_1+\vec{x}_2$,  i.e., the \emph{number of representations}. 
For this purpose, one considers a fixed $\vec{e}_2$ of weight $v$ and computes the number of $\vec{x_1}$ of weight  $v'$, such that  $\vec{e_2}-\vec{x_1}$ is of weight $v'$.
Let us denote this number of representations by $r$ and $u= \log_q(r)$.

Let us now recall how one performs a representation merge: 
given two lists $\mathcal{L}_1, \mathcal{L}_2$ containing $\vec{x}_i$ of weight $v'$, we add $\vec{x}=\vec{x}_1+\vec{x}_2$ to the resulting list $\mathcal{L}$, whenever $\vec{x}$ attains a target weight $v$ and $\vec{x}\mat{A}_2^\top = \vec t$, on the first $u$ positions, for either $\vec t=\vec{s}_2$, the target syndrome or $\vec t=\vec 0$, the zero vector. 
As for any $\vec{x} \in \mathcal{L}$ there are $r$ representations $(\vec{x}_1, \vec{x}_2)$, which all lead to the same $\vec{x}$, by checking on $u$ positions, it is guaranteed that one representation for each possible $\vec{x}$ survives the merge with high probability. 
In general, a representation merge of two lists $\mathcal{L}_1, \mathcal{L}_2$ of asymptotic size $\Lambda$ on $u$ positions costs asymptotically
$$
2^{n\max\{\Lambda, 2\Lambda-U\}}.
$$
After the representation merge, one performs a filtering step, which removes vectors which are not well-formed, e.g., do not achieve a given weight constraint.
Further steps can then utilize this smaller list. 

In the following, we denote by BJMM$(a)$ an algorithm that starts with a concatenation merge followed by $a$ representation merges, since the optimal number of levels $a$ might change depending on the parameters.

For a BJMM algorithm with $a$ levels,  we denote by $\Sigma_i$ the size of the search space, by $V_i = \lim\limits_{n \to \infty} \frac{v_i(n)}{n}$, with $V_0=V$, the weight of the vectors, by $E_i= \lim\limits_{n \to \infty}\frac{\varepsilon_i(n)}{n}$ the number of overlaps on level $i \in \{0, \ldots, a\}$, starting from $a$. With $U_{i}=Q\cdot\lim\limits_{n \to \infty}\frac{u_{i}(n)}{n}$ we denote the number of positions on which we merge.
  \begin{theorem}
The enumeration cost of the BJMM(2) algorithm  is given by
\begin{align}\label{eq:bjmm_enum_cost}
\max\{ \tfrac{\Ls_2}{2}, \Ls_2-U_1, 2\Ls_2-U_1-U_0,   2\Ls_1-U_0-QL\} , 
\end{align}  
where we can optimize $V_i,E_i$ under the constraints that
\begin{align*} 
\Pwt_{i} &= \frac{\Pwt_{i-1}}{2} + \Ewt_i, \text{ and }
\Ls_i = (R+L)\hbin{\tfrac{V_i}{R+L}}+ZV_i, \\
U_i &=V_i+(R+L-V_i)\hbin{\tfrac{E_{i+1}}{R+L-V_i}}+ZE_{i+1}.
\end{align*}
\end{theorem}
This algorithm can be used with any number of levels $a$, and the cost of a restricted BJMM($a$) algorithm follows straightforwardly. 
Furthermore, for the subsequent modifications, we always refer to the cost \eqref{eq:bjmm_enum_cost}, which can be computed using only the sizes of the search spaces and the number of representations. 

In the following, we present some new algorithms derived from BJMM to solve the RSDP. 
For small choices of $z$, we can take advantage of the structure of $\mathbb{E}.$ 
Following the idea of  \cite{bcj}, we add a few  elements, denoted by $\mathbb{E}_+,$ from $\{\alpha+\beta \mid \alpha, \beta \in \mathbb{E}\}$ to the restricted set $ \mathbb{E}$, to have more representations.  
Since we added new elements to the search space, in the final representation merge these elements in $\mathbb{E}_+$ need to add up to elements in $\mathbb{E}_0.$
We call those algorithms BJMM$(a)_+$, to denote also the number of levels $a$.

\subsection{Case $z=2$}\label{subsec:z2}

We generalize the classical BJMM-like approach by allowing $1+1=2$ (and $-1 + (-1) = -2$) in intermediate lists. 
We call this the $\mathrm{BJMM}(2)_+$ algorithm. 
Thus, in this case we have $\mathbb{E}= \{ \pm 1\}$ and $\mathbb{E}_+= \{\pm 2\}.$ 
This changes the number of representations.
In order to construct the intermediate lists using representation merge, we have the usual $v_i$ entries in $\{\pm 1 \}$ and we also require $m_i$ to denote the number of $\pm2$'s on level $i$.
Then, the number of well-formed vectors on level $i$ is given by $\binom{k+l}{v_i, m_i} 2^{v_i+m_i}$.

\begin{figure}
\centering
\begin{tikzpicture}
\def\shifta{0.85cm}
\def\shiftb{1.0cm}
\node[color = black] at (-.5,0.25){$\vec{e}^{(i)\hphantom{+1}}$};
\draw[pattern=north west lines] (0,0) rectangle (3,0.5); 
\node[] at (1.5,-.25){$V_i$};
\draw[pattern=crosshatch dots] (3,0) rectangle (5,0.5); 
\node[] at (4,-.25){$M_i$};
\draw[] (0,0) rectangle (7,0.5);

\node[color = black] at (-.5,0.25cm+\shiftb){$\vec{e}^{(i+1)}_2$};
\draw[] (0,\shiftb) rectangle (1.5,0.5cm+\shiftb);
\draw[pattern=north west lines] (1.5,\shiftb) rectangle (3,0.5cm+\shiftb); 
\draw[] (0,\shiftb) rectangle (7,0.5cm+\shiftb);
\draw[pattern=crosshatch dots] (0.5,\shiftb) rectangle (1,0.5cm+\shiftb); 
\draw[pattern=north west lines] (5.5,\shiftb) rectangle (6.25,0.5cm+\shiftb);
\draw[pattern=crosshatch dots] (3.5,\shiftb) rectangle (4.0,0.5cm+\shiftb);
\draw[pattern=north west lines] (4,\shiftb) rectangle (5,0.5cm+\shiftb);

\node[] at (-.675,0.25cm+\shiftb+\shifta/2){$+$};

\node[color = black] at (-.675,0.75cm){\rotatebox{90}{$=$}};
\node[color = black] at (-.5,0.25cm+\shiftb+\shifta){$\vec{e}^{(i+1)}_1 $};
\node[] at (.75,.75cm+\shiftb+\shifta){$V_i/2$};
\draw[pattern=north west lines] (0,\shiftb+\shifta) rectangle (1.5,0.5cm+\shiftb+\shifta);
\draw[] (0,\shiftb+\shifta) rectangle (7,0.5cm+\shiftb+\shifta);
\node[] at (2.25,.75cm+\shiftb+\shifta){$B_{i+1}$};
\draw[pattern=crosshatch dots] (2,\shiftb+\shifta) rectangle (2.5,0.5cm+\shiftb+\shifta);
\node[] at (5.875,.75cm+\shiftb+\shifta){$E_{i+1}$};
\draw[pattern=north west lines] (5.5,\shiftb+\shifta) rectangle (6.25,0.5cm+\shiftb+\shifta); 
\draw[pattern=crosshatch dots] (3,\shiftb+\shifta) rectangle (3.5,0.5cm+\shiftb+\shifta); 
\node[] at (4.5,.75cm+\shiftb+\shifta){$C_{i+1}$};
\draw[pattern=north west lines] (4,\shiftb+\shifta) rectangle (5,0.5cm+\shiftb+\shifta); 
\end{tikzpicture}
\caption{Counting the number of representations for level $i$.}
\label{fig:num_repr24}
\end{figure}
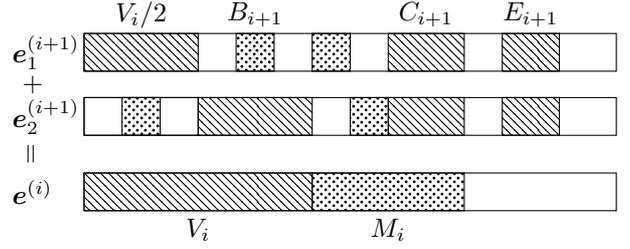

The number of representations of $\vec{e}^{(i)}=\vec{e}^{(i+1)}_1+ \vec{e}^{(i+1)}_2$ on level $i$ is counted as per Figure \ref{fig:num_repr24}. For this it is enough to count the number of $\vec{e}^{(i+1)}_1$. There are $\binom{v_i}{v_i/2}$ ways of splitting the support of the elements in $\Rset$, without choosing the entries.  
Out of the chosen $v_i/2$  we chose $b_{i+1}$ positions, that  overlap with $\pm 2$'s in $\vec{e}^{(i+1)}_2$ and also in the $v_i/2$ non-chosen positions we choose $b_{i+1}$ many positions to be $\pm 2$. For this we have $\binom{v_i/2}{b_{i+1}}^2$ possibilities. 
Out of the $m_i$ many $\pm2$'s on level $i$, $c_{i+1}$ are constructed as $\pm(1+1)$.
The remaining $(m_i-c_{i+1})$ many $\pm2$'s are obtained by support splitting.
This results in $\binom{m_i}{(m_i-c_{i+1})/2,c_{i+1}}$. 
Finally, one can choose $\varepsilon_{i+1}$ out of the $k+\ell-m_i-v_i$ zero-positions.   

While this enlarges the number of well-formed vectors on the intermediate levels, the number of representations is also increased.
Let 
$M_i = \lim\limits_{n \to \infty} \frac{m_i(n)}{n}, B_i = \lim\limits_{n \to \infty} \frac{b_i(n)}{n}, C_i = \lim\limits_{n \to \infty} \frac{c_i(n)}{n}$. 
Then, the following corollary holds.

\begin{corollary}\label{cor:BJMMplus_z2}
The exponent of the enumeration cost of the $\mathrm{BJMM}(2)_+$ algorithm is calculated according to \eqref{eq:bjmm_enum_cost}.
Using
\begin{align*}
V_{i} = \frac{V_{i-1}}{2} + E_i + C_i    \text{ and }
M_{i} = \frac{M_{i-1}-C_i}{2} + B_i,    
\end{align*}
results in
\begin{align*}
\Ls_i & =  (R+L)\trin{\tfrac{\Pwt_{i}}{R+L}}{\tfrac{M_{i}}{R+L}} +  Z (\Pwt_i+M_i), \\
\Repr_i  & = 
\Pwt_i\left(1+\hbin{{\tfrac{2B_{i+1}}{\Pwt_i}}}\right) +
M_i  \trin{\tfrac{C_{i+1}}{M_i}}{\tfrac{M_i - C_{i+1}}{2M_i}}\\
 & +(  R+L-\Pwt_i-M_i)\hbin{\tfrac{\Ewt_{i+1}}{R+L-\Pwt_i-M_i}} +  Z \Ewt_{i+1}.
\end{align*}
\end{corollary}

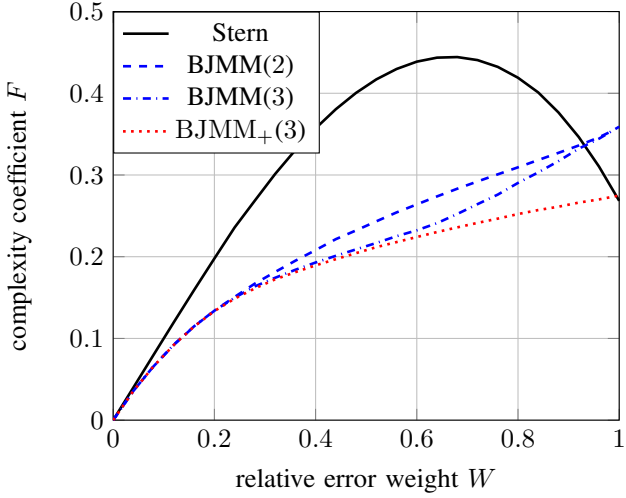
\begin{figure}[ht!]
\centering
\begin{tikzpicture}
\begin{axis}[
width = .93\columnwidth,
height = 7cm,
xmin =  0,
xmax= 1,
grid = major,
ymin=0,
ymax=0.5,
legend style={at={(0,1)},anchor=north west},
xlabel={relative error weight $W$},
ylabel={complexity coefficient $F$},
]
\addplot[black, line width=1pt]coordinates{
( 0      , 0      )
( 0.0400 , 0.0400 ) 
( 0.0800 , 0.0801 ) 
( 0.1200 , 0.1200 ) 
( 0.1600 , 0.1593 ) 
( 0.2000 , 0.1978 ) 
( 0.2400 , 0.2358 ) 
( 0.2800 , 0.2687 ) 
( 0.3200 , 0.3012 ) 
( 0.3600 , 0.3305 ) 
( 0.4000 , 0.3571 ) 
( 0.4400 , 0.3805 ) 
( 0.4800 , 0.4005 ) 
( 0.5200 , 0.4170 )
( 0.5600 , 0.4297 )
( 0.6000 , 0.4387 ) 
( 0.6400 , 0.4436 ) 
( 0.6800 , 0.4443 ) 
( 0.7200 , 0.4406 )
( 0.7600 , 0.4323 ) 
( 0.8000 , 0.4190 ) 
( 0.8400 , 0.4010 )
( 0.8800 , 0.3765 ) 
( 0.9200 , 0.3465 ) 
( 0.9600 , 0.3104 ) 
(1,0.2684)
};\addlegendentry{Stern};

\addplot[blue,dashed, line width=1pt]coordinates{
( 0      , 0      )
( 0.0400 , 0.0358 ) 
( 0.0800 , 0.0660 ) 
( 0.1200 , 0.0922 ) 
( 0.1600 , 0.1147 )
( 0.2000 , 0.1339 )
( 0.2400 , 0.1508 ) 
( 0.2800 , 0.1671 ) 
( 0.3200 , 0.1813 )
( 0.3600 , 0.1951 ) 
( 0.4000 , 0.2081 )
( 0.4400 , 0.2215 ) 
( 0.4800 , 0.2322 ) 
( 0.5200 , 0.2433 )
( 0.5600 , 0.2542 ) 
( 0.6000 , 0.2640 )
( 0.6400 , 0.2738 ) 
( 0.6800 , 0.2832 )
( 0.7200 , 0.2921 ) 
( 0.7600 , 0.3008 )
( 0.8000 , 0.3094 ) 
( 0.8400 , 0.3183 ) 
( 0.8800 , 0.3271 ) 
( 0.9200 , 0.3361 ) 
( 0.9600 , 0.3461 ) 
( 0.9900 , 0.3551 )
};\addlegendentry{BJMM($2$)};

\addplot[blue, dashdotted, line width=1pt]coordinates{
( 0      , 0      )
( 0.0400 , 0.0359 ) 
( 0.0800 , 0.0661 ) 
( 0.1200 , 0.0926 ) 
( 0.1600 , 0.1149 ) 
( 0.2000 , 0.1340 ) 
( 0.2400 , 0.1498 ) 
( 0.2800 , 0.1637 ) 
( 0.3200 , 0.1742 ) 
( 0.3600 , 0.1840 ) 
( 0.4000 , 0.1929 )
( 0.4400 , 0.2011 ) 
( 0.4800 , 0.2090 ) 
( 0.5200 , 0.2168 ) 
( 0.5600 , 0.2251 ) 
( 0.6000 , 0.2323 ) 
( 0.6400 , 0.2411 ) 
( 0.6800 , 0.2523 ) 
( 0.7200 , 0.2643 ) 
( 0.7600 , 0.2760 ) 
( 0.8000 , 0.2903 )
( 0.8400 , 0.3032 ) 
( 0.8800 , 0.3170 ) 
( 0.9200 , 0.3322 ) 
( 0.9600 , 0.3444 ) 
( 1.0000 , 0.3591 ) 
};\addlegendentry{BJMM($3$)};

\addplot[red, dotted, line width=1pt]coordinates{
( 0      , 0      )
( 0.0400 , 0.0358 )
( 0.0800 , 0.0660 ) 
( 0.1200 , 0.0921 ) 
( 0.1600 , 0.1146 ) 
( 0.2000 , 0.1335 ) 
( 0.2400 , 0.1487 ) 
( 0.2800 , 0.1612 ) 
( 0.3200 , 0.1723 ) 
( 0.3600 , 0.1811 ) 
( 0.4000 , 0.1894 ) 
( 0.4400 , 0.1970 ) 
( 0.4800 , 0.2044 ) 
( 0.5200 , 0.2117 ) 
( 0.5600 , 0.2179 ) 
( 0.6000 , 0.2241 ) 
( 0.6400 , 0.2302 ) 
( 0.6800 , 0.2359 )
( 0.7200 , 0.2414 )
( 0.7600 , 0.2466 ) 
( 0.8000 , 0.2522 ) 
( 0.8400 , 0.2571 )
( 0.8800 , 0.2612 ) 
( 0.9200 , 0.2662 )
( 0.9600 , 0.2701 ) 
( 1.0000 , 0.2749 ) 
};\addlegendentry{$\mathrm{BJMM}_+$($3$)};
\end{axis}
\end{tikzpicture}
\caption{
Comparison of the complexity coefficients $F(R,W)$ for restricted Stern/Dumer,
the general adaption of BJMM and the generalization given in Corollary \ref{cor:BJMMplus_z2} using $q=157$, $z=2$ and $R=0.5$.}\label{fig:T_plot_z2}
\end{figure}

Figure \ref{fig:T_plot_z2} shows the curve of the complexity coefficient $F(R,W)$ for $R=0.5$, $q=157$ and $z=2$.
Classical ISD algorithms are usually compared by going through all rates and fixing the weight as large as possible under the uniqueness condition, as this results in the hardest instances. 
However, for the new RSDP, this is not true in general. 
Thus, we chose to fix $R=0.5$ and go through all weights $W \in [0,1]$, as they all allow for unique decoding. 
It can be observed that the adapted BJMM algorithm improves significantly over restricted Stern/Dumer for medium error weights.
While for the classical SDP two representation levels give the best performance \cite{bjmm}, here, three representation layers were found to be optimal.
The generalization given in Corollary \ref{cor:BJMMplus_z2} provides a further improvement for increasing error weights.
It was observed that the number of elements from $\Rset_{+}$ is optimized to approximately $0$ in the base lists. 
Hence, one can start with restricted base lists and not lose a noticeable amount of performance.

In \cite{oldrest}, the case of $z=2$ is considered with the particular choice of $W=1$.
As can be seen from Figure \ref{fig:T_plot_z2}, in this weight regime, the approach of Corollary \ref{cor:BJMMplus_z2} does not offer any improvement over Stern/Dumer.
For such instances, it is advantageous to shift the error vector $\vec{e}$ to $\tilde{\vec{e}}=\left(\vec{e}+(1, \ldots, 1)\right)/2 \in\{0,1\}^n$, which is done by computing  $\tilde{\vec s} = \left(\vec s + (1, \ldots, 1)\mat H^\top\right)/2$, see e.g. \cite{ternary}. 
The resulting error weight is approximately $\tilde{w} = n/2$.
In order to solve the transformed instance, we follow the BCJ approach \cite{bcj}, i.e., increase the number of representations by allowing $-1$'s in intermediate lists, which have to be added to the base lists. In the level $i$, for a $\vec{e}^{(i)} \in \{ 0, \pm1\}^{(k+\ell)}$  with $v_{i}$ $1$'s and $m_{i}$ $-1$'s we  write $\vec{e}^{(i)}=\vec{e}_1^{(i+1)}+\vec{e}^{(i+1)}_2$ as in Figure \ref{fig:num_repr24}, with $b_{i+1}=c_{i+1}=0$ and outside of the support we choose $\varepsilon_{i+1}$ $1$'s that cancel with $-1$ and $\varepsilon_{i+1}$ $-1$'s that cancel with $1$'s. 
\begin{corollary}
The exponent of the shifted BCJ(2) algorithm is calculated according to  \eqref{eq:bjmm_enum_cost} with
\begin{align*}
V_{i} &=  \frac{V_{i-1}}{2} + E_i \text{ and }
M_{i} = \frac{M_{i-1}}{2} + E_i, \\ 
    \Ls_i &= (R+L)\cdot\trin{\frac{V_i}{R+L}}{\frac{M_i}{R+L}},\\
    U_i &= M_i + V_i\\
    &+ (R+L-V_i-M_i)\hbin{\frac{2E_{i+1}}{R+L-V_i-M_i}} + 2E_{i+1}. 
\end{align*}
\end{corollary}

\subsection{Case $z=4$}\label{subsec:z4}
In this case, we have $\mathbb{E}= \{ \pm 1, \pm g\}$ and define $\mathbb{E}_+= \{ \pm (g + 1), \pm (g-1)\}.$
Following the approach for $z=2$, we obtain the same number of possibilities for choosing the supports of $\vec{e}^{(i+1)}_1$ and $\vec{e}^{(i+1)}_2$.
There are, however, more possibilities for picking the values in the chosen positions:
there are  two possibilities for obtaining any $e\in\Rset$ as the sum of $\alpha\in\Rset$ and $\beta\in\Rset_+$ and two possibilities for obtaining $e\in\Rset_+$ as the sum of two elements in $\mathbb{E}.$ 
This increases the number of representations for level $i$ overall by factor of $2^{2(b_{i+1}+c_{i+1})}$ compared to the number computed for $z=2$.
Hence, we obtain the same $V_i, M_i, \Ls_i$ as in Corollary \ref{cor:BJMMplus_z2} and  the  new number of representations is given by
\begin{align*}
\Repr_i  & = 
\Pwt_i\left(1+\hbin{{\tfrac{2B_{i+1}}{\Pwt_i}}}\right) +
M_i  \trin{\tfrac{C_{i+1}}{M_i}}{\tfrac{M_i - C_{i+1}}{2M_i}}\\
 & +(  R+L-\Pwt_i-M_i)\hbin{\tfrac{\Ewt_{i+1}}{R+L-\Pwt_i-M_i}} +  Z \Ewt_{i+1} \\ &  + 2(B_{i+1}+C_{i+1}).
\end{align*}

\begin{figure}[h!]
\centering
\begin{tikzpicture}
\begin{axis}[
xmin =  0,
xmax= 1,
grid = major,
ymin=0,
ymax=0.8,
width = .93\columnwidth,
height = 7cm,
legend style={at={(0,1)},anchor=north west},
xlabel={relative error weight $W$},
ylabel={complexity coefficient $F$},
]
\addplot[black, line width=1pt]coordinates{
( 0      , 0      )
( 0.0400 , 0.0406 ) 
( 0.0800 , 0.0832 ) 
( 0.1200 , 0.1271 ) 
( 0.1600 , 0.1705 ) 
( 0.2000 , 0.2162 ) 
( 0.2400 , 0.2626 ) 
( 0.2800 , 0.3091 )
( 0.3200 , 0.3552 ) 
( 0.3600 , 0.4033 ) 
( 0.4000 , 0.4463 ) 
( 0.4400 , 0.4851 ) 
( 0.4800 , 0.5237 ) 
( 0.5200 , 0.5592 ) 
( 0.5600 , 0.5913 ) 
( 0.6000 , 0.6195 ) 
( 0.6400 , 0.6437 ) 
( 0.6800 , 0.6632 ) 
( 0.7200 , 0.6779 ) 
( 0.7600 , 0.6871 ) 
( 0.8000 , 0.6904 )
( 0.8400 , 0.6868 ) 
( 0.8800 , 0.6756 ) 
( 0.9200 , 0.6553 ) 
( 0.9600 , 0.6239 ) 
( 1.0000 , 0.5795 ) 
};\addlegendentry{Stern};

\addplot[blue,dashed, line width=1pt]coordinates{
( 0      , 0      )
( 0.0400 , 0.0382 ) 
( 0.0800 , 0.0741 ) 
( 0.1200 , 0.1084 ) 
( 0.1600 , 0.1411 ) 
( 0.2000 , 0.1725 ) 
( 0.2400 , 0.2025 ) 
( 0.2800 , 0.2317 ) 
( 0.3200 , 0.2590 ) 
( 0.3600 , 0.2855 ) 
( 0.4000 , 0.3116 ) 
( 0.4400 , 0.3362 ) 
( 0.4800 , 0.3615 ) 
( 0.5200 , 0.3866 ) 
( 0.5600 , 0.4108 ) 
( 0.6000 , 0.4370 ) 
( 0.6400 , 0.4653 ) 
( 0.6800 , 0.4955 ) 
( 0.7200 , 0.5262 ) 
( 0.7600 , 0.5569 ) 
( 0.8000 , 0.5878 ) 
( 0.8400 , 0.6189 )
( 0.8800 , 0.6500 )
( 0.9200 , 0.6814 ) 
( 0.9600 , 0.7128 )
( 1.0000 , 0.7444 ) 
};\addlegendentry{BJMM($2$)};

\addplot[red, dashdotted, line width=1pt]coordinates{
( 0      , 0      )
( 0.0400 , 0.0382 ) 
( 0.0800 , 0.0740 ) 
( 0.1200 , 0.1082 ) 
( 0.1600 , 0.1408 ) 
( 0.2000 , 0.1719 )
( 0.2400 , 0.2016 ) 
( 0.2800 , 0.2297 )
( 0.3200 , 0.2566 ) 
( 0.4000 , 0.3057 ) 
( 0.4400 , 0.3283 )
( 0.4800 , 0.3500 )
( 0.5200 , 0.3711 ) 
( 0.5600 , 0.3916 ) 
( 0.6000 , 0.4168 ) 
( 0.6400 , 0.4310 ) 
( 0.6800 , 0.4499 ) 
( 0.7200 , 0.4683 ) 
( 0.7600 , 0.4864 ) 
( 0.8000 , 0.5039 ) 
( 0.8400 , 0.5210 ) 
( 0.8800 , 0.5377 )
( 0.9200 , 0.5541 )
( 0.9600 , 0.5704 ) 
( 1.0000 , 0.5865 ) 
};\addlegendentry{$\mathrm{BJMM}_+$($2$)}; 

\addplot[red, dotted, line width=1pt]coordinates{
( 0      , 0      )
( 0.0400 , 0.0382 )
( 0.0800 , 0.0741 ) 
( 0.1200 , 0.1083 ) 
( 0.1600 , 0.1411 ) 
( 0.2000 , 0.1725 ) 
( 0.2400 , 0.2018 ) 
( 0.2800 , 0.2300 ) 
( 0.3200 , 0.2570 ) 
( 0.3600 , 0.2851 ) 
( 0.4000 , 0.3060 ) 
( 0.4400 , 0.3331 ) 
( 0.4800 , 0.3499 ) 
( 0.5200 , 0.3700 ) 
( 0.5600 , 0.3895 )
( 0.6000 , 0.4078 ) 
( 0.6400 , 0.4239 ) 
( 0.6800 , 0.4402 ) 
( 0.7200 , 0.4556 ) 
( 0.7600 , 0.4704 ) 
( 0.8000 , 0.4847 ) 
( 0.8400 , 0.4985 ) 
( 0.8800 , 0.5116 )
( 0.9200 , 0.5243 ) 
( 0.9600 , 0.5372 )
( 1.0000 , 0.5500 ) 
};\addlegendentry{$\mathrm{BJMM}_+$($3$)}; 
\end{axis}
\end{tikzpicture}
\caption{
Comparison of the complexity coefficients $F(R,W)$ for restricted Stern/Dumer,
the general adaption of BJMM and the proposed generalization using $q=157$, $z=4$ and $R=0.5$.}\label{fig:T_plot_z4}
\end{figure}
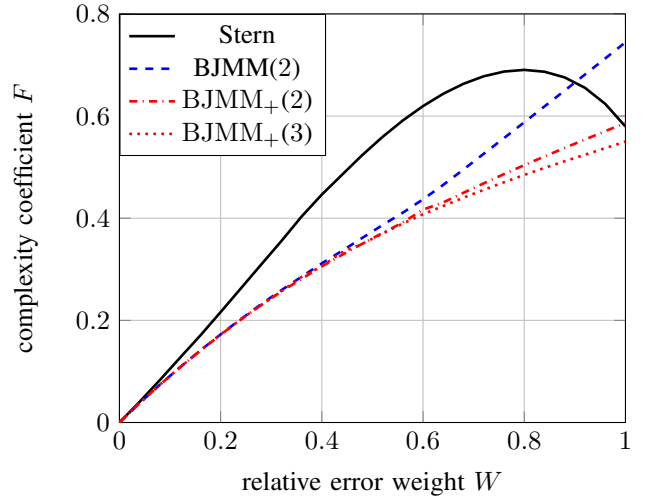

Figure \ref{fig:T_plot_z4} shows the curve of the complexity coefficient $F(R,W)$ for $R=0.5$, $q=157$ and $z=4$.
Again, BJMM improves over Stern for medium error weights.
The generalization using $\Rset_+$ gives a further speedup for increased weights.

\subsection{Case $z=6$}\label{subsec:z6}
In this case, we have $\mathbb{E}=\{ \pm1, \pm g, \pm(g-1)\}$. 
Note that $\mathbb{E}$ already possesses additive structure: 
 any element $e\in\Rset$ can be obtained as $e = e_1 + e_2 = e_2 +e_1$ with $e_1,e_2\in\Rset, e_1\neq e_2$.
This allows setting $\Rset_+= \Rset$.
Thus, using again Figure \ref{fig:num_repr24}, there are $\binom{v_i}{2b_{i+1}, v_i/2-b_{i+1}} 2^{2b_{i+1}}$ representations due to the entries of $\vec{e}^{(i+1)}_1$ within the $v_i$ positions of $\vec{e}^{(i)}$ in $\mathbb{E}$.
The remaining factors are calculated as before, setting $m_{i}=c_i=0$.

\begin{corollary}\label{cor:BJMMplus_z6}
The exponent of the enumeration cost of the $\mathrm{BJMM}(2)_+$ algorithm is calculated according to \eqref{eq:bjmm_enum_cost} using
\begin{align*} 
V_{i} &=  \tfrac{V_{i-1}}{2} + E_i + B_i, \\
\Ls_i &=   (R+L)\hbin{\tfrac{V_{i}}{R+L}} +  Z V_i, \\
\Repr_i  &=   \Pwt_i\trin{\tfrac{2B_{i+1}}{\Pwt_i}}{\tfrac{\Pwt_i/2 - B_{i+1}}{\Pwt_i}}+ 2B_{i+1}\\ & 
+(R+L-\Pwt_i) \hbin{ \tfrac{\Ewt_{i+1}}{R+L-\Pwt_i}} +  Z \Ewt_{i+1}.
\end{align*}
\end{corollary}

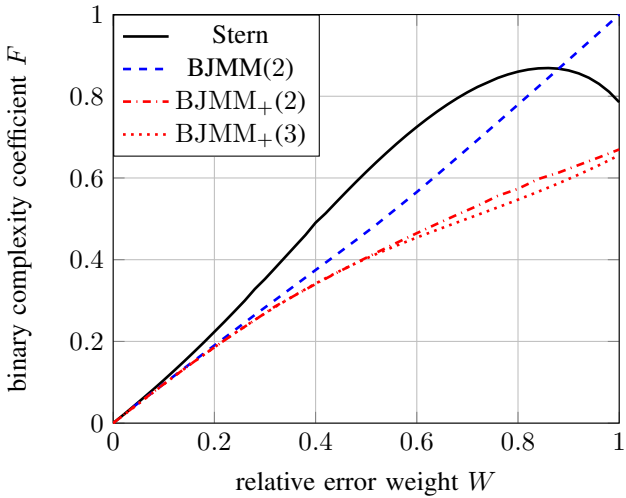
\begin{figure}[h!]
\centering

\begin{tikzpicture}
  \begin{axis}[
xmin =  0,
xmax= 1,
grid = major,
ymin=0,
ymax=1.0,
width = .93\columnwidth,
height = 7cm,
legend style={at={(0,1)},anchor=north west},
xlabel={relative error weight $W$},
ylabel={binary complexity coefficient $F$},
]
\addplot[black, line width=1pt]coordinates{
(0,0)
( 0.0200 , 0.0202 )
( 0.0400 , 0.0409 ) 
( 0.0600 , 0.0620 ) 
( 0.0800 , 0.0834 ) 
( 0.1000 , 0.1054 ) 
( 0.1200 , 0.1280 ) 
( 0.1400 , 0.1510 ) 
( 0.1600 , 0.1746 )
( 0.1800 , 0.1986 ) 
( 0.2200 , 0.2483 )
( 0.2400 , 0.2739 ) 
( 0.2600 , 0.2998 )
( 0.2800 , 0.3283 ) 
( 0.3000 , 0.3528 ) 
( 0.3200 , 0.3796 ) 
( 0.3400 , 0.4066 ) 
( 0.3600 , 0.4336 ) 
( 0.3800 , 0.4605 ) 
( 0.4000 , 0.4906 ) 
( 0.4200 , 0.5137 ) 
( 0.4400 , 0.5397 ) 
( 0.4600 , 0.5653 ) 
( 0.4800 , 0.5904 ) 
( 0.5000 , 0.6148 )
( 0.5200 , 0.6385 ) 
( 0.5400 , 0.6614 ) 
( 0.5600 , 0.6836 ) 
( 0.5800 , 0.7048 ) 
( 0.6000 , 0.7250 ) 
( 0.6200 , 0.7442 ) 
( 0.6400 , 0.7624 )
( 0.6600 , 0.7794 ) 
( 0.6800 , 0.7952 ) 
( 0.7000 , 0.8097 ) 
( 0.7200 , 0.8228 ) 
( 0.7600 , 0.8448 ) 
( 0.7800 , 0.8533 ) 
( 0.8000 , 0.8602 ) 
( 0.8200 , 0.8652 ) 
( 0.8400 , 0.8682 ) 
( 0.8600 , 0.8690 )
( 0.8800 , 0.8675 ) 
( 0.9000 , 0.8632 )
( 0.9200 , 0.8559 ) 
( 0.9400 , 0.8452 ) 
( 0.9600 , 0.8303 )
( 0.9800 , 0.8109 )
( 1.0000 , 0.7855 )
};\addlegendentry{Stern};

\addplot[dashed, blue, line width=1pt]coordinates{
( 0.0400 , 0.0400 ) 
( 0.0800 , 0.0774 ) 
( 0.1200 , 0.1152 ) 
( 0.1600 , 0.1528 ) 
( 0.2000 , 0.1902 ) 
( 0.2400 , 0.2274 ) 
( 0.2800 , 0.2650 )
( 0.3200 , 0.3013 ) 
( 0.3600 , 0.3379 )
( 0.4000 , 0.3747 ) 
( 0.4400 , 0.4112 ) 
( 0.4800 , 0.4481 ) 
( 0.5200 , 0.4857 ) 
( 0.5600 , 0.5256 ) 
( 0.6000 , 0.5656 )
( 0.6400 , 0.6075 ) 
( 0.6800 , 0.6501 )
( 0.7200 , 0.6929 ) 
( 0.7600 , 0.7361 ) 
( 0.8000 , 0.7796 ) 
( 0.8400 , 0.8232 ) 
( 0.8800 , 0.8671 )
( 0.9200 , 0.9112 )
( 0.9600 , 0.9555 ) 
( 1.0000 , 1.0000 ) 
};\addlegendentry{BJMM($2$)};

\addplot[red, dashdotted, line width=1pt]coordinates{
( 0,       0)
( 0.0400 , 0.0389 ) 
( 0.0800 , 0.0770 ) 
( 0.1200 , 0.1141 ) 
( 0.1600 , 0.1506 ) 
( 0.2000 , 0.1856 ) 
( 0.2400 , 0.2197 ) 
( 0.2800 , 0.2525 ) 
( 0.3200 , 0.2857 ) 
( 0.3600 , 0.3133 ) 
( 0.4000 , 0.3405 ) 
( 0.4400 , 0.3667 ) 
( 0.4800 , 0.3923 ) 
( 0.5200 , 0.4171 )
( 0.5600 , 0.4413 ) 
( 0.6000 , 0.4649 ) 
( 0.6400 , 0.4878 ) 
( 0.6800 , 0.5103 )
( 0.7200 , 0.5319 ) 
( 0.7600 , 0.5562 ) 
( 0.8000 , 0.5735 ) 
( 0.8400 , 0.5974 ) 
( 0.8800 , 0.6130 ) 
( 0.9200 , 0.6318 ) 
( 0.9600 , 0.6504 ) 
( 1.0000 , 0.6698 ) 
};\addlegendentry{$\mathrm{BJMM}_+$($2$)};

\addplot[red, dotted, line width=1pt]coordinates{
( 0,       0)
( 0.1200 , 0.1141 ) 
( 0.1600 , 0.1504 ) 
( 0.2000 , 0.1857 ) 
( 0.2400 , 0.2199 ) 
( 0.2800 , 0.2528 ) 
( 0.3200 , 0.2842 ) 
( 0.3600 , 0.3138 ) 
( 0.4000 , 0.3417 ) 
( 0.4400 , 0.3676 ) 
( 0.4800 , 0.3916 ) 
( 0.5200 , 0.4139 ) 
( 0.5600 , 0.4346 ) 
( 0.6000 , 0.4542 ) 
( 0.6400 , 0.4730 ) 
( 0.6800 , 0.4914 ) 
( 0.7200 , 0.5096 ) 
( 0.7600 , 0.5281 ) 
( 0.8000 , 0.5469 ) 
( 0.8400 , 0.5664 ) 
( 0.8800 , 0.5868 ) 
( 0.9200 , 0.6083 ) 
( 0.9600 , 0.6312 ) 
( 1.0000 , 0.6557 )
};\addlegendentry{$\mathrm{BJMM}_+$($3$)}; 
\end{axis}
\end{tikzpicture}
\caption{
Comparison of the complexity coefficients $F(R,W)$ for restricted Stern/Dumer,
the general adaption of BJMM and the generalization given in Corollary \ref{cor:BJMMplus_z6} using $q=157$, $z=6$ and $R=0.5$.}\label{fig:T_plot_z6}
\end{figure}

Figure \ref{fig:T_plot_z6} shows the curve of the complexity coefficient $F(R,W)$ for $R=0.5$, $q=157$ and $z=6$.
Using the additive structure of $\Rset$ as proposed in Corollary \ref{cor:BJMMplus_z6} enables a remarkable speedup over Stern/Dumer and the basic BJMM adaption.
Experiments, for which we allowed elements from $\left\{\pm(g+1),\pm(2g-1),\pm(g-2)\right\}$ in intermediate lists, did not yield further performance improvements.

\subsection{Security Level Update for Instances from Literature}\label{sec:params}

In the following, we apply the presented algorithms to the parameters proposed in \cite{freudenberger,oldrest,oldrest2,freudenberger2}.
We did not perform a rigorous finite regime analysis, since already approximating the security level as $2^{F(R,W)\cdot n}$ shows a considerable reduction, compared to the original analysis. 
The obtained results are summarized in Table \ref{Tab:params}.
\urlstyle{tt}
Python code for reproducing the work factors and the parameters of the decoders is publicly available at \url{github.com/sebastianbitzer/rest-dec}.

\begin{table}[h!] 
\centering
\caption{Estimates of the work factors of the proposed algorithms compared to claimed security levels for instances from literature.}\label{Tab:params}

\begin{tabular}{
  @{}
  S[table-format=3.0, table-column-width=0.3cm,  table-number-alignment=left]
  S[table-format=1.0, table-column-width=0.3cm,  table-number-alignment=center] 
  S[table-format=5.0, table-column-width=0.78cm, table-number-alignment=center] 
  S[table-format=3.0, table-column-width=0.45cm, table-number-alignment=center]
  S[table-format=1.2, table-column-width=0.5cm,  table-number-alignment=center] 
  S[table-format=1.2, table-column-width=0.5cm,  table-number-alignment=center] 
  S[table-format=3.0, table-column-width=0.6cm,  table-number-alignment=center]
  S[table-format=3.0, table-column-width=0.9cm,  table-number-alignment=right] 
  S[table-format=3.0, table-column-width=1.0cm,  table-number-alignment=right] 
  @{}
}
\toprule
       &
 {$z$} &
 {$q$} &
 {$n$} &
 {$R$} &
 {$W$} &
 {claim} &
 \multicolumn{2}{c}{this work}\\
 &
 &
 &
 &
 &
 &
 {(\si{\bit})}&
 \multicolumn{2}{c}{{(\si{\bit})}} \\
\midrule
\cite{oldrest2}       & 2 & 16381 & 400 & 0.75 & 0.16 & 128& {$\mathrm{BJMM}(2)_+$} & 69\\ 
\cite{oldrest2}       & 2 & 16381 & 500 & 0.75 & 0.13 & 128& {$\mathrm{BJMM}(2)_+$} & 76\\ 
\cite{oldrest2}       & 2 & 16381 & 400 & 0.80 & 0.14 & 128& {$\mathrm{BJMM}(2)_+$} & 69\\ 
\cite{oldrest2}       & 2 & 32749 & 500 & 0.75 & 0.13 & 128& {$\mathrm{BJMM}(2)_+$} & 76\\ 
\cite{oldrest2}       & 2 & 32749 & 600 & 0.66 & 0.14 & 128& {$\mathrm{BJMM}(2)_+$} & 82\\ 
\cite{oldrest}        & 2 & 29 & 167 & 0.79 & 1.00 &  87 & {shifted $\mathrm{BCJ}(3)$}&48\\ 
\cite{oldrest}        & 2 & 31 & 256 & 0.80 & 1.00 & 128 & {shifted $\mathrm{BCJ}(3)$}&74\\ 
\cite{freudenberger}  & 4 & 109 & 270 & 0.50 & 0.34 & 125 &{$\mathrm{BJMM}(2)_+$} &75\\ 
\cite{freudenberger}  & 4 & 157 & 312 & 0.50 & 0.34 & 144 &{$\mathrm{BJMM}(2)_+$} &85 \\ 
\cite{freudenberger}  & 4 & 197 & 384 & 0.50 & 0.34 & 177 & {$\mathrm{BJMM}(2)_+$}&103\\ 
\cite{freudenberger2} & 4 & 137 & 272 & 0.20 & 0.60 &  88 &{$\mathrm{BJMM}(2)_+$} &45\\ 
\cite{freudenberger2} & 4 & 157 & 312 & 0.20 & 0.60 & 101 &{$\mathrm{BJMM}(2)_+$} &51\\
\cite{freudenberger2} & 4 & 173 & 344 & 0.20 & 0.60 & 111 &{$\mathrm{BJMM}(2)_+$} &56\\
\cite{freudenberger2} & 4 & 193 & 384 & 0.20 & 0.60 & 124 &{$\mathrm{BJMM}(2)_+$} &62\\ 
\cite{freudenberger}  & 6 & 139 & 276 & 0.02 & 0.60 &  89 &{$\mathrm{BJMM}(3)_+$} &50 \\
\cite{freudenberger}  & 6 & 157 & 312 & 0.20 & 0.60 & 101 &{$\mathrm{BJMM}(3)_+$} &55 \\ 
\cite{freudenberger}  & 6 & 193 & 384 & 0.20 & 0.60 & 124 &{$\mathrm{BJMM}(3)_+$} &67 \\ 
\bottomrule
\end{tabular}
\end{table}

\subsection{Arbitrary $z$}\label{subsec:z_general}
While the presented attacks focused on error sets of size $z\in\{2,4,6\}$, the proposed solvers can also be generalized to larger values of $z$. 
Such a generalization can utilize any additive structure of $\Rset_0$.
The concrete structure of $\Rset_0$ depends on the factorization of $x^z-1$ in $\mathbb{F}_q$. 
As the factorization cannot be given in general, any proposed $z$ should be checked independently.
As we have seen, it can be beneficial to allow elements from $\mathbb{E}_+$ to increase the number of representations.
Finally, the possibility of transforming the problem by shifting the error vector has to be taken into account.

 \section{Conclusion}
 \label{sec:conclusion}
In this paper, we studied the complexity of the restricted syndrome decoding problem, which has recently gained attention in code-based cryptography.
To this end, we adapted the representation technique, which is utilized in the fastest known solvers for syndrome decoding problems to the new setting.
In particular, small choices of the restriction cardinality $z$ were considered, for which we provided novel tailored solvers which are inspired by~\cite{bcj}. 
This leads to a drastic decrease of the respective security levels.  
Nevertheless, we believe that the restricted syndrome decoding problem is a promising underlying problem for cryptographic applications.
In contrast to previous proposals, we would like to advocate the use of larger values of $z$.
 
 \section*{Acknowledgements}
Sebastian Bitzer acknowledges the financial support by the Federal Ministry of Education and Research of Germany in the programme of ``Souverän. Digital. Vernetzt.''. Joint project 6G-life, project identification number: 16KISK002. Violetta Weger  is  supported by  the European Union's Horizon 2020 research and innovation programme under the Marie Sk\l{}odowska-Curie grant agreement no. 899987.
\IEEEtriggeratref{12}

\bibliographystyle{plain}
\bibliography{references}
 
\end{document}